\documentstyle[epsfig,prd,aps,12pt]{revtex}
\newcommand\lsim{\mathrel{\rlap{\lower4pt\hbox{\hskip1pt$\sim$}}      
\raise1pt\hbox{$<$}}}
\newcommand\gsim{\mathrel{\rlap{\lower4pt\hbox{\hskip1pt$\sim$}}      
\raise1pt\hbox{$>$}}} 
\begin{document}
\draft
%@twocolumnfalse\endcsname
\preprint{DF-FCUL/CFNUL-99/05}
\title{Slow-roll inflation without  fine-tuning
\thanks{preprint} }
\author{Tiago C. Charters$^{1,2}$
\thanks{email:charters@alf1.cii.fc.ul.pt}, Jos\'e P. Mimoso$^{1,2}$\thanks{email:jpmimoso@alf1.cii.fc.ul.pt} \&
Ana Nunes$^{1,3}$\thanks{email: anunes@lmc.fc.ul.pt}}
\address{$^{1}$Dep. F\'{\i}sica, F.~C.~L., Ed C1, Campo Grande, 1700
Lisboa, Portugal}
\address{$^2$CFNUL and $^3$CMAF, Av. Prof. Gama Pinto, 2, 1699 Lisboa Codex, Portugal}
\date{\today}
\maketitle
  
\begin{abstract}
The slow-roll approximation  is the usual starting  point to study the
constraints  imposed on   the  inflaton potential  parameters   by the
observational data. We  show that, for a potential  exhibiting at least two extrema
and giving rise to a limited inflationary  period, slow-roll does  not have to be taken
as an additional hypothesis  and is in fact  forced by the constraints
on the number of e-foldings.  
\end{abstract}

\section{Introduction}
Inflation avoids the horizon,  flatness and monopole problems of the
standard cosmological   model  by just  requiring  a   brief period of
accelerated expansion during the very early universe~\cite{K+T 90}. It also provides
an explanation for the origin of the  fluctuations associated with the
observed  cosmic microwave background  anisotropies~\cite{LL93}. The latter is a
most valued feature at present, since after the detection made by COBE
of these anisotropies the  prospects  for probing  the physics of  the
inflationary  models became real.  Moreover the  coming MAP and Planck
missions will push our ability to contrast the theoretical models with
observations to  the precision required to rule  out many of  them.   

The underlying  ideas   of inflation are well-kown, but a fully
consistent scenario is still  unavailable. In most cases the predictions of
the  models  are   based   on   a   few  features  that    seem
consensual. In the simplest version of inflation one ordinarily
considers a self-interacting  scalar field $\varphi$ (the {\em inflaton})
being damped by expansion of the  universe and eventually yielding the negative
pressures  required  to promote  the violation   of the  strong energy
principle,  and  hence the positive   acceleration.  The details of the
process  depend on the form of the potential,  but the popular
approach has been to associate the bulk of inflation with
a stage during which the scalar field slow rolls along a potential well.
The {\em slow-roll parameters} are defined in this context and a
perturbative expansion  in  terms of these parameters~\cite{Lidsey et al 97,LPB 94}
is then used to predict the anisotropies of the cosmic microwave background (CMB), and
conversely, to reconstruct from the latter data the shape of the inflaton potential.

A central issue regarding this usual procedure is then to ascertain the legitimacy
of the slow-roll approximation.
Given an arbitrary potential the slow-roll regime is not a generic
dynamical pattern. However  the vast majority of inflationary models rely on the slow-roll
assumption and then show that it is possible to constrain the parameters of the potential in
such a way as to meet the requirements of enough inflation and limited fluctuations in
the CMB~\cite{LL93}.

The issues of the genericity of the slow-roll regime and of the possibility of meeting
the observational requirements without the slow-roll assumptions have not until now been
dealt with satisfactorily. In what concerns the former of these questions, there has been
some effort to either avoid it or argue in favour of an affirmative answer in the case
of particular models~\cite{Linde 83,M+C 88,K+B 89,Freese et al 90,Adams et al 93,G+P 92,K+O 93}. 
In particular the results of Ref.~\cite{K+O 93} have shown that for
the pseudo-Nambu-Goldstone-boson of the natural inflation model~\cite{Freese et al 90}
fine tuning can be avoided if we consider non-zero initial velocities for the field.
We recover below this conclusion in an independent context.
As for the latter question there has not been to our knowledge any attempt
to address it directly. Leaving this question unanswered encourages criticism of
the relevance of the theoretical estimates based on the slow-roll approximation.
Recently, the reliability of the reconstruction procedure based on
the lowest-order slow-roll expansion~\cite{Lidsey et al 97} has
been questioned by Wang, Mukhanov and Steinhardt~\cite{Wang et al  98}
who show that its validity depends
on the effective equation of state of the inflaton field being almost stationary
during the relevant epoch of horizon exit of the fluctuations.    
In this reference the authors explicitly consider a potential for which
this condition is not met and show that the implications for the CMB
anisotropies are significantly different from those predicted by the
slow-roll approximation. In Ref.~\cite{Copeland et al 98} it is shown that the
particular potential considered in~\cite{Wang et al  98} is such that several requirements,
among which the end of inflation, cannot be met. 

Given that slow-roll is an atypical dynamical regime for damped oscillations, in the
absence  of a general negative answer to the second of the questions, the
validity of the slow-roll approximation will remain a natural target to criticism such as
in~\cite{Wang et al  98}. 

In the present work our aim has been to determine the consequences of imposing the
constraints on the number of e-foldings and on the amplitude of the CMB fluctuations to
a general scalar field model giving rise to a limited inflationary period.
Our study relies on a numerical analysis of the
dynamics, which is both rigorous and general because it stems from the knowledge of the
global dynamics of the non-linear dynamical system under consideration.
Contrary to what may be thought the numerical approach is more adequate than
an alternative analytic approach which is bound here to be either local or too
restrictive.

The results of this analysis show that the  constraints on the  number of
e-foldings force the free parameters  that control the basic features
of the potential to take the values  for which the orbits exhibit slow
roll dynamics for a relatively long  time of their evolution.  This is
a consequence of the known  fact that  rapid oscillations in  convex
potentials  do not  contribute   to
inflation~\cite{Turner 83,Damour+Mukhanov 98,Liddle+Mazumdar 98}.
We also show that these values of the  parameters
correspond to a big region in parameter space. These two conclusions are of great
importance on the one hand to avoid the criticism of~\cite{Wang et al  98} and, on the other
hand, to establish that inflation is likely to happen.

We  consider  the flat isotropic Friedmann-Robertson-Walker  (FRW)
universe characterized by the metric 
\begin{equation}
ds^2=-dt^2  +  a(t)^2  \left[dr^2  +  r^2(d\theta^2  +  \sin{\theta}^2
d\phi^2)\right]\quad , 
\end{equation}
and assume that  the matter content of the  universe is  a homogeneous
scalar field with a  self-interacting potential.  The field  equations
are 
\begin{equation}
H^2 =  {8\pi \over 3 {m^2_{pl}}} \left({1\over 2} {\dot{\varphi}}^2 +
V(\varphi)\right) \; , \label{eq_Fried}
\end{equation}
\begin{equation}
\ddot\varphi + 3H\dot\varphi + V'(\varphi)= 0 \; , \label{eq_sf}
\end{equation}
where the overdots  stand for the  derivatives with respect to  time, $m^2_{pl}$
is the Planck mass, $H= \dot{a}/a$ is the Hubble parameter,
and $V'={\rm d}V/{\rm d}\varphi$.
It is also possible to deduce from the latter equations that
\begin{equation}
\frac{\ddot a}{a} ={8\pi \over 3 {m^2_{pl}}}\left(
V(\varphi)-\dot\varphi^2\right) 
\end{equation}
holds and hence that inflation requires $V(\varphi)>\dot\varphi^2$.

In our numerical study, we shall consider potentials of the form 
\begin{eqnarray}
V_1(A,B,\varphi) &=& \frac{A}{B^2} (\varphi^2-B)^2 \; , \label{pot_DW} \\ 
V_2(A,B,\varphi) &=&  A\, \left(1+\cos(\frac{\pi\varphi}{\sqrt{B}})\right) \; , \label{pot_PNGB}
\end{eqnarray}
where $A$  and $B$   are  positive constants. However,
our results extend to potentials exhibiting several extrema. For these potentials, given any
initial conditions outside a potential trap, the damping term in
Eq.~(\ref{eq_sf}) will make the system evolve to approach the potential
well where it will be caught by the stable equilibrium. Provided that
this is vanishing (no false vacuum term), the last stage of the
evolution of the system corresponds to a potential with one maximum and
one vanishing minimum, such as in Eqs.~(\ref{pot_DW},\ref{pot_PNGB}), together
with initial conditions slightly above the energy of the maximum.
Futhermore, the behaviour of the solutions of the systems associated
with the potentials $V_1$ of Eq.~(\ref{pot_DW}) or $V_2$ of
Eq.~(\ref{pot_PNGB}) is typical in this more general class: the
qualitative behaviour dos not depend on the detailed form of the
potential and is described in some detail below. In this general setting our results apply to
any non-degenerate potential with the aforementioned qualitative properties, since the only
essential requirement is the non-degeneracy of the consecutive extrema.
>From a physical viewpoint, our conclusions
hold for a non-degenerate  potential susceptible of giving rise to a finite inflationary period
whose end is followed by the death of the oscillating field, and also to sufficient inflation
without fine-tuning the parameters. In fact, the first condition excludes potentials without
extrema, such as the exponential potential or the arctan potential of Ref.~\cite{Wang et al  98},
as well as potentials with a false vacuum term~\cite{GB+Wands 96,Kinney 97},
while the second discards single minimum potentials as the quadratic
potential~\cite{Linde 83,M+C 88,Liddle+Mazumdar 98}.

Consider the Eqs.~(\ref{eq_Fried},\ref{eq_sf}) with the potentials
(\ref{pot_DW}) or (\ref{pot_PNGB}). In both  cases and indeed for any
potential with a maximum and a vanishing minimum, the problem reduces to the
study of the dynamics  of a non-linear, non-linearly damped,
oscillator whose damping term tends to zero as the stable equilibrium is approached.
In spite of the high  non-linearity,  the equations are those of an
autonomous  system in  the plane,  and   therefore the possibility  of
chaotic behavior is excluded. Indeed, the qualitative description
of the behavior of the system in phase-space is quite straightforward.
Consider for instance the case of the standard double well potential: there
are three equilibrium points,  one saddle at $(\varphi,\dot\varphi) =
(0,0)$ and  two   stable  focus  at $(\varphi,\dot\varphi)  =   (\pm
\sqrt{B},0)$.  The stable manifolds of  the saddle point separate the
plane into two regions, I and  II. One of  the stable focuses attracts
all the  points in I,   and the other  attracts all  of the points in 
II. Moreover, there   is  no
divergence of nearby orbits, and, apart from the  boundary of the two
basins of attraction, the future behaviour of the orbits will change smoothly
with the initial conditions. Bearing this picture in mind,  it is possible
to perform a thorough  numerical exploration of  the system,  in order to
determine the  set   of parameters  and initial   conditions  which  satisfy
the observational constraints. 

It is usual to consider the parameters~\cite{LL93,LPB 94}
\begin{equation}
\epsilon(t)    =  \frac{3}{2} \frac{\dot{\varphi}^2}{ \frac{\dot{\varphi}^2}{2} +
V(A,B,\varphi)} 
\end{equation}
\begin{equation}
\eta(t) = - \frac{\ddot{\varphi}}{H \dot{\varphi}} \ .
\end{equation}
Inflation is equivalent to having $\epsilon <1$.
The slow roll regime is defined by the following conditions: $\epsilon
(t) \ll  1$,which  is  the condition for   neglecting  the kinetic energy contribution
to  the damping term in Eq.~(\ref{eq_sf}),  and $\eta (t) \ll   1$,
which  is  the  condition for neglecting  the acceleration term in
Eq.~(\ref{eq_sf}). 

To succesfully solve the cosmological puzzles of standard cosmology an
inflationary model must satisfy the following constraints~\cite{K+T 90,LL93}. 

{\em{(i) Sufficient  inflation}} - We demand that  the scale factor of
the universe inflates by at least 60 e-foldings.

{\em{(ii) Energy  bounds}}  - To keep full generality we allow inflation to
take place bewteen the Planck energy  scale of $m_{pl}$ and the electroweak
phase transition at $10^{-17}m_{pl}$.  This provides energy bounds in phase-space for the
dynamics of the scalar field $ 10^{-34} m^2_{pl}\le H^2 \le m^2_{pl}$.

{\em{(iii)    Density perturbations}} -  The observations of the anisotropies of the CMB
constraint the amplitude of density fluctuations, $\delta_H$, to satisfy the bound
$\delta_H\le \delta^{obs}_H \simeq 1.9\times 10^{-5}$~\cite{COBE,B+L+W 96}.

We shall now impose these constraints on the dynamics of Eqs.~(\ref{eq_Fried},\ref{eq_sf})
under the potentials $V_i(A,B,\varphi)$, $i=1,2$. 
First   note that  both   potentials are
proportional to $A$. Changing the time scale through $t = \tau/\sqrt{A}$ and
redefining the field's velocity and acceleration accordingly, the equations of
motion~(\ref{eq_Fried},\ref{eq_sf}) become independent of the parameter $A$. 
The total number of e-foldings is given by
\begin{equation}
N(A,B)   =     \int_0^{+\infty}    \sqrt{{8\pi  \over    3  {m^2_{pl}}}
\left(\frac{{\dot{\varphi}}^2(t)}{2}+V(A,B,\varphi(t))\right)} dt  \; ,
\end{equation}
where $\varphi(t)$ is a solution of the Eqs.~(\ref{eq_Fried},\ref{eq_sf}).
Performing the rescaling of the time variable in the latter integral, we
have $N(A,B) = N(1,B)$, and, thus, the number of e-foldings is independent of $A$. 
Notice that this scaling property is also independent of the detailed
form of $V$.

When $\epsilon \ll 1$ and $\dot{\epsilon} \simeq 0$ hold, it is known that
$\delta_H= {\cal{O}}(1)H^2 /\dot{\varphi}$, where the right-hand side is
evaluated at horizon crossing~\cite{LL93,Wang et al 98}.  Thus, assuming
that $\epsilon(t)$ satisfies the
latter conditions,  the amplitude of the  density  fluctuations 
scales  as   $\delta_H(A)    =      \sqrt{A}\,
\delta_H(1)$ and so restriction (iii) can be dealt with separately.
Setting $\varphi(0) =0$, we are thus left with the two
parameters $(B,\varphi'(0))$, where the prime denotes
the derivative with respect to the rescaled time, and we have to satisfy
the former two constraints, (i) and (ii).

\begin{figure}
\center
\epsfig{file=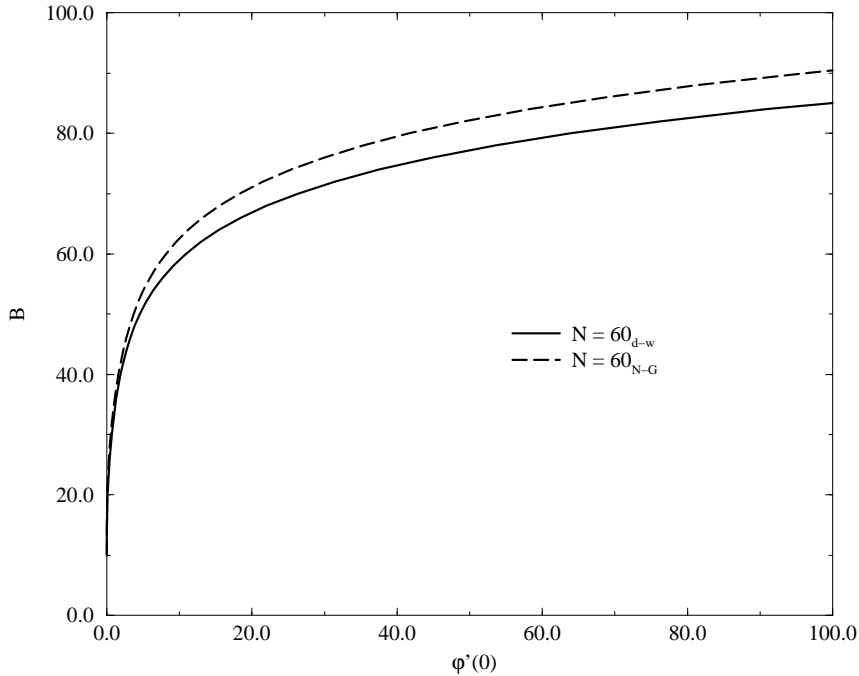,height=13cm,angle=-90}
\caption{\small{In this figure we represent the boundary of the admissible region
in parameter space from the viewpoint of the constraints on the number of e-foldings for
the  double-well potential (labelled d-w) and the pseudo-Nambu-Goldstone-boson potential of
natural inflation (N-G). The initial energy bounds translate in terms of $\varphi'(0)$  in
an upper bound of the order of $10^5$.}}
\label{fig1}
\end{figure}

The initial energy bounds define an interval of admissible values for $\varphi'(0)$.
The usual approach is to study the ensemble of initial conditions with
energy  equal or lower than the symmetry breaking  energy $A$.  Here we
consider initial energies that are equal or larger than $A$. This is important
to avoid fine tuning, since these orbits will have an additional contribution to the
number of e-foldings with respect to the lower energy orbits usually considered
in the literature.

The question then reduces to find for each $\varphi'(0)$ in the admissible energy range
the interval of values of $B$ such that  $N(1,B)\in [60,+\infty]$. The results are shown
in Figure~1, and it can be seen that the constraint on the number of e-foldings can be met
in a big region of parameter space. More interestingly, it turns out that for all
the parameter values that satisfy the constraint on the number of e-foldings,
the slow-roll parameters $\epsilon$ and $\eta$, as well as the time derivatives
of $\epsilon$,  behave in such a way that the slow-roll assumptions are fulfilled
during the whole inflationary period. Thus the slow-roll approximation  is valid.
In Figure~2 we represent a typical orbit in the admissible region in parameter space, as well
as the behaviour of $\epsilon$  and $\eta$. In Figure~3 we show the typical values of $\epsilon$
and $\eta$ for those orbits during the inflationary period. This means that the slow-roll
assumption to deal with the constraint on $\delta_H$ is fully justified. Moreover, the fact
that slow-roll holds during the inflationary period agrees with the observational constraints
on the spectral index and on the gravitational waves contribution to the
CMB~\cite{LL93,A+F 95,G+L 96,G+L 99,Zibin 99a,Zibin 99b}.

More significantly,
our analysis provides a negative answer to the question of whether it is possible to meet the
basic inflationary requirements without slow-roll in the class of
potentials with several extrema yielding a finite inflationary period.
In fact, our numerical results amount to a global study of 
the $(B,\dot \varphi(0))$ parameter  plane   and show
that parameters that give rise to orbits off the slow-roll regime also
fail to satisfy the constraint on the number of e-foldings.
This finding agrees  with the results of \cite{Copeland et al 98} regarding the
potential considered  in~\cite{Wang et al  98},  tailor-made
to   break  up  the    slow-roll
approximation, and which   they   show gives  rise to   a   non-flat
perturbation spectrum. The idea that the slow-roll regime is forced by
the observational constraints seems to  prevail in the general setting
of  arbitrary potentials exhibiting symmetry breaking, or just several extrema,
as well as for other specific  models.

It is  also  clear that the three  main
constraints,  on  the number  of e-foldings,  on the amplitude  of  the
perturbation spectrum, and on the initial  energy of the field, can be
met in a potential belonging to the broad class considered in this paper without
fine tuning of the parameters.  The
final conclusion is then that in this setting the
slow-roll approximation is a general  starting point to study further 
constraints imposed by new observational data.  
\begin{figure}
\center
\epsfig{file=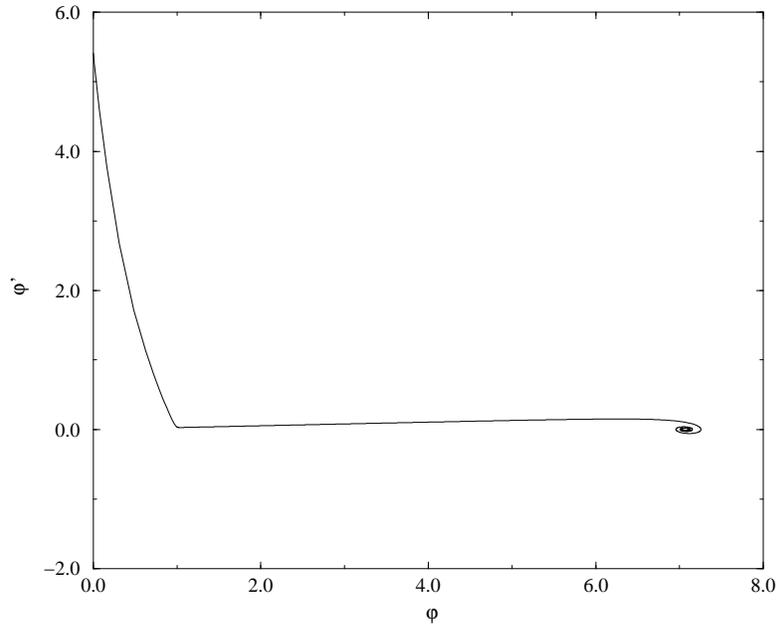,height=12cm,angle=-90}
\epsfig{file=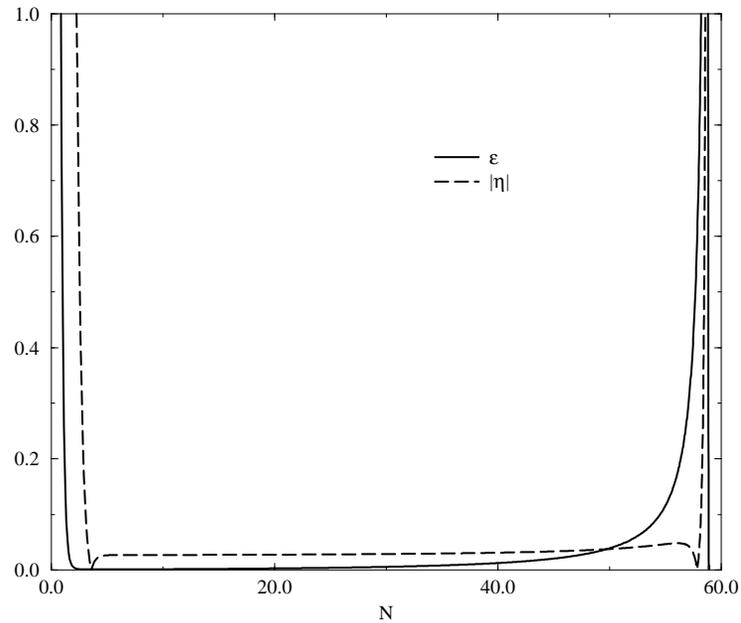,height=12cm,angle=-90}
\caption{\small{In figure (a) we plot a typical slow-roll orbit in phase-space with
$B=50$ and initial conditions $\varphi(0)=0$ and $\varphi'(0)=5.41533$ for the double-well
potential. The PNGB is analogous. In (b) we show the behaviour of the slow-roll parameters
along the orbit in (a).} }
\label{fig2}
\end{figure}

\begin{figure}
\center
\epsfig{file=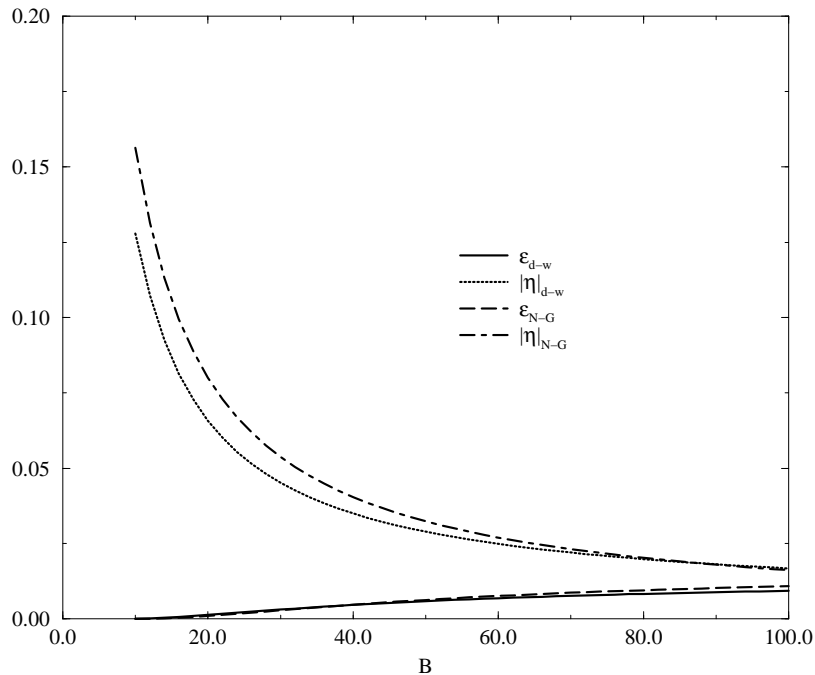,height=13cm,angle=-90}
\caption{\small{Slow-roll parameters as a function of $B$ evaluated at $N=30$.} }
\label{fig3}
\end{figure}
%%%%%%%%%%%%%%%%%%%%%%
\section*{Acknowledgements}
The authors wish  to  acknowledge the  finantial support  from Funda\c c\~ao  de
Ci\^encia e Tecnologia   under the grant PBIC/C/FIS/2215/95 and
A.N. thanks the project PRAXIS 2/2.1/MAT/125/94. We are also grateful
to Andrew R. Liddle for helpful comments on an preliminary draft of this work. 
%%%%%%%%%%%%%%%%%%%%%%
%%%%%%%%%%%%%%

%%%%%%%%%%%%%%%%%%%%%%
%%%%%%%%%%%%%%%%%%%%%%
%%%%%%%%%%%%%%%%%%%%%%
\end{document}